\documentclass[preprint,aps,epsfig,prb]{revtex4}
\usepackage{graphicx}

\begin{document}
\title{Comparison of a Microgel Simulation to Poisson-Boltzmann Theory
   \footnote{For submission to the Journal of Chemical Physics}}

\author{Gil C. Claudio, Christian Holm, and Kurt Kremer}

\affiliation{Max-Planck-Institut f\"ur Polymerforschung,
             Ackermannweg 10,
             55128 Mainz,
             Germany}

\begin{abstract}

We have investigated a single charged microgel in aqueous solution
with a combined simulational model and Poisson-Boltzmann theory.
In the simulations we use a coarse-grained charged bead-spring model
in a dielectric continuum, with explicit counterions and full electrostatic
interactions under periodic and non-periodic boundary conditions.
The Poisson-Boltzmann model is that of a single charged colloid  
confined to a spherical cell where the counterions are allowed to  
enter the uniformly charged sphere. We compare the simulational results
to those of the Poisson-Boltzmann solution and find good agreement, i.e.,
for the number of confined counterions within the gel. We then proceed
to investigate the origin of the differences between the results
these two models give, and performed a variety of simulations
which were designed to test for the influence of charge correlations,
excluded volume interactions, and thermal fluctuations in the strands
of the gel.  Our results support the applicability of the Poisson-Boltzman
cell model to study ionic properties of small microgels under dilute conditions.

\end{abstract}

\maketitle

\section{Introduction}

Hydrogel research has constantly attracted a lot of attention,
due to its capability of absorbing large quantities of water,
resulting in the expansion of the material up to 1000 times its dry volume.
Recent review articles on hydrogel research range from general
reviews\cite{Hoare_Kohane_2008} to application specific reviews such as
sensors\cite{Richter_etal_2008}, surface applications\cite{Netz_Anderlman_2003},
drug delivery applications\cite{Oh_Matyjaszewski_etal_2008}
and other biomedical applications\cite{Klouda_Mikos_2008,Park_etal_2007,Baroli_2007}.

A hydrogel is a crosslinked network of water soluble polymers.
The hydrophilicity of the free uncrosslinked chain allows it
to attract water molecules and be soluble in water up to infinite
dilution.  The crosslinking, however, restricts the dilution to a
maximal swelling of the gel.  The combination of these two properties
allows the hydrogel to attract water molecules and keep them within
the network, resulting in an overall swelling of the entire gel.
Hydrogels can be made of neutral or charged polymers called polyelectrolytes
(the backbone of polyelectrolytes, however, need not be hydrophilic).
When polyelectrolyte gels are dissolved in water, charges are formed
in the polyelectrolyte when  the counter ions dissociate from the polymer.
These counter ions are free to move within the solvent, but stay, in the
absence of salt, within the vicinity  of the polyelectrolyte to keep
electrostatic neutrality of the gel.  The increase in entropy due to these
counter ions overcomes the extra enthalpy to bring in even more water
molecules into the network. The overall effect is that charged hydrogels
have an even greater swelling ratio than their uncharged counterparts.
A hydrogel matrix can be made to a few centimeters in diameter, depending
on the synthetic procedure used.  Microgels, on the other hand, are hydrogels
that are within the micrometer range.

A number of analytical theories on hydrogels and polyelectrolytes have been
presented\cite{Joanny_Barrat_1996,Homl_Joanny_Kremer_etal_2004,Mann_etal_2004,Schiessel_Pincus_1998,Schiessel_1999}.
Corresponding simulations have also been
done\cite{Mann_etal_2005,Yan_dePablo_2003,Schneider_Linse_2002},
confirming the basic picture that the entropy of the counter ions exerts
an extra osmotic pressure which in turn increases the swelling capacity
of the hydrogels.  Thus determining the location of these counter ions within
the network, and specifically the investigation of  ion condensation around
the polymer chains, is an important aspect of hydrogel research.
To our knowledge, the simulations that have been done, however,
all consider quasi infinite systems via periodic boundary conditions that
connect the chains of one side of the simulation box to its opposite side.
This set up suffers from the limitation that
the counter ions are, in a sense, always inside the polymer matrix.
Thus they cannot examine the actual charge fraction of counter ions
inside the network which could, in turn, change the amount of osmotic pressure
exerted by the counter ions, and thus change the swelling properties.

In an attempt to attack this problem, we simulate a model microgel, which is a
finite sized object made up of crosslinked polymers in a periodic box.
The simulation details are described in section \ref{sec_mg_sim_details}.
The system is a spherical polymer network in a simulation box,
as shown in figure \ref{fig_microgel}.

This model simulation allows us to calculate the counter ion condensation not only
within the vicinity of each chain, but also the fraction inside the sphere, and 
study the consequences for the chain conformations.  This model is also similar to an
even more reduced model of a single charged sphere with counter ions that can be 
situated inside or oustide this charged sphere.  This reduced model can be calculated
using Poisson-Boltzmann (PB) theory.  The simulation results can therefore be compared
to a PB calculation, and thus provide evidence for the applicability of the Poisson-Boltzman
cell model to study ionic properties of microgels.

There exists, however, quite a number of differences between the
Poisson-Boltzmann method and simulation set-up, which can potentially confuse
the interpretation of a direct comparison between the two. This paper examines
the effects of these differences by presenting model simulations aimed at systematically
removing, or adding, these factors in a step-wise fashion, with the final aim of
studying the usefulness of the PB theory for a description of the swelling properties
of microgels.

In section \ref{sec_mg_pb_methodology}, we describe the details of the 
microgel (MG) simulation (\ref{sec_mg_sim_details}), brief description of PB theory
(\ref{sec_pb_description}), and lastly the differences between PB calculation and the
microgel simulation (\ref{sec_pb_to_mg}).  This list of differences (or assumptions)
provides the rationale for the rest of the model simulations that we performed,
aimed at connecting, in a stepwise manner, the PB model to the MG simulation.
The methodologies and results of these simulations described in sections \ref{sec_ls}
through \ref{sec_ssmg}. And in section \ref{sec_conc}, we give our counclusions.

\section{Microgel Simulation and Poisson-Boltzmann Calculation Methodologies}
\label{sec_mg_pb_methodology}

In all simulations, the following reduced units were used: energy is placed
in terms of $k_BT$ which is given a value of $1.0$, values of lengths are
in $\sigma$, charge is in units of $e$, and time is in $\tau$.  All particles
are given a mass of $1.0$.
Table \ref{tab_sim_common} summarizes the parameters that are the same
in all simulations in this paper, whereas table \ref{tab_param_pb_ls_st}
and \ref{tab_param_pb_mg}
lists the parameters that are specific to each simulation.  All simulations were done
using ESPResSo\cite{Espresso_2006}, a molecular dynamics package
for soft matter simulations, under NVT conditions.  In order to simulate a constant
temperature ensemble, a Langevin thermostat is used, where all particles are coupled
to a heat bath.  Thus the equation of motion is modified to
\begin{equation}
m_i {d^2 \vec{r}_i \over dt^2 } = - \vec{\nabla}_i V_{tot} \left( \vec{r}_i \right) \\
- m_i \Gamma { d \vec{r}_i \over dt } + \vec{W}_i \left( t \right)
\end{equation}
where $m_i$ is the mass of particle $i$, $V_{tot}$ is the total potential acting
on the particle,
$\Gamma$ is the friction coefficient that is set to $1.0$, and $\vec{W}_i(t)$ 
is a stochastic force that is uncorrelated in time and across particles.  The friction
and stochastic forces are linked by the fluctuation-dissipation theorem
\begin{equation}
\left<\vec{W}_i\left(t\right) \cdot \vec{W}_j\left(t\right)\right> = \\
6 m \Gamma k_B T \delta_{ij} \delta\left(t-t'\right)
\end{equation}
where $k_B T=1.0$.

\subsection{Microgel Simulation Details}
\label{sec_mg_sim_details}
The parameters for the microgel simulation are listed in Tables \ref{tab_sim_common}
and \ref{tab_param_pb_mg}, labelled as \textbf{P-MG}, where \textbf{P} indicates
that periodic boundary conditions were used. Figure \ref{fig_microgel}a shows
how the starting structure of the microgel was constructed, whereas
figure \ref{fig_microgel}b gives a snapshot of an equilibrated microgel configuration. 
The microgel was simulated using a general bead-spring model \cite{Grest_Kremer_1986,Kremer_Grest_1990}.
The microgel was composed of 46 polymers, each composed of 10 monomers (beads),
giving a total of 460 monomers. These polymers were connected by 29 crosslinks,
each represented by one bead, giving a total of 489 beads (monomers plus crosslinks)
in the microgel.  Some crosslinks, especially those inside the microgel, were connected
to the ends of four different polymers (serving as tetrafunctional nodes),
while the rest were conneced to the ends of three different polymers (trifunctional
nodes).  Since all of the polymer ends were connected to one crosslink,
the microgel had no dangling chains.  Upon examining figure \ref{fig_microgel}a,
one could see that the starting structure resembles that of a diamond lattice,
with the edges cut off to approach a spherical shape.  Once the particles are 
allowed to move, the average shape further approximates that of a sphere.
The whole set-up was chosen to have an average equilibrated shperical
shape, in order to make a comparison with the PB calculation of colloid using
a cell model.

Half of the monomers (that is, 230) were given a charge of +1.0, while the others
remained uncharged.  The charged and uncharged monomers were placed along the polymer
in an alternating sequence.  Out of the 29 total crosslinks, 20 were also
given a charge of +1.0, resulting to a total of 250 of the 489 total particles
in the microgel having a charge of +1.0.  Thus 250 free counter ions,
each with a charge of -1.0, were added to the simulation,
giving a total 739 objects and an overall neutral charge.

The non-bonded interactions of the particles were described by a Weeks-Chandler-Anderson
potential \cite{WCA_Potential_1971} of the form
\begin{equation}
\label{eq_lj}
U_{LJ} (r_{ij}) = \left\{
\begin{array}{l l}
4 \epsilon_{LJ} \left[ \left( \sigma_0 \over r_{ij} \right)^{12} - 
\left( \sigma_0 \over r_{ij} \right)^6 - c_{shift} \right] \qquad \mbox{for $r_{ij} < r_{cut}$} \\
0 \qquad \qquad \qquad \qquad \qquad \qquad \qquad \mbox{for $r_{ij} \geq r_{cut}$}
\end{array} \right.
\end{equation}
where $\sigma_0 = 1.0 $, $\epsilon_{LJ} = 1.0$, $r_{cut} = 2^{1/6}$,
and $c_{shift} = 0.25$, thus leaving only the repulsive
part of the interaction.  These are the standard parameters for bead-spring model
of polymers in a good solvent \cite{WCA_Potential_1971}.
The bonds between the particles were modelled using the FENE (Finite Extension
Nonlinear Elastic) potential\cite{Grest_Kremer_1986}
\begin{equation}
\label{eq_fene}
U_{FENE} (r) = \left\{
\begin{array}{l l}
- \frac{1}{2} k_F r_F^2 \ln \left[ 1 - \left( r \over r_F \right)^2 \right] \qquad \mbox{for $r < r_F$} \\
\infty \qquad \qquad \qquad \qquad \qquad \mbox{for $r \geq r_F$}
\end{array} \right.
\end{equation}
where $k_F=10.0 \frac{k_BT}{\sigma^2}$ and $r_F=1.5$ \cite{Yan_dePablo_2003}.
These parameters allow the bonds to equlibrate to $r\simeq1.0$.

The potential energy between charges $q_i$ and $q_j$ separated by a distance $r_{ij}$
were described by the unscreened Coulomb potential
\begin{equation}
\label{eq_coulomb}
U_C (r_{ij}) = {k_B T l_B q_i q_j \over r_{ij}}
\end{equation}
where $l_B$, the Bjerrum length, is defined as
\begin{equation}
\label{eq_bjerrum}
l_B = {e_0^2 \over {4 \pi \epsilon_0 \epsilon_r k_B T}}
\end{equation}
where $e_0$ is the elementary charge, $\epsilon_0$ is the vacuum permittivity, and $\epsilon_r$
is the relative dielectric constant of the solvent.  The Bjerrum length
is defined as the distance at which the Coulomb energy between two unit
charges is equal to the interaction energy of $k_BT$.  This parameter
provides a convenient way of tuning the strength of electrostatic interactions in
a simulation.  As a point of reference, the Bjerrum length of water at room 
temperature is $7.1$ \AA.  In this paper, a Bjerrum length of $l_B=1$
and $k_BT=1$ was used in all simulations (Table \ref{tab_sim_common}).
Given a typical hydrogel such as poly(N-isopropylacrylamide) (poly-NIPAAm), the
average distance between the centers of masses between two repeat units is within the
order of $7.1$ \AA, making the choice of $r=1.0$ and $l_B=1.0$ both
computational convenient and comparable to values of actual hydrogels.

The electrostatics for this simulation box was treated using the P3M
algorithm\cite{Deserno_Holm_1998a,Deserno_Holm_1998b}, tuned to an accuracy
of $10^{-4}$ in the absolute error of the Coulomb forces
with the help of the formulas in Ref \cite{Deserno_Holm_1998b}.

The MG simulation was performed in a cubic box with a length of $84$ and
periodic boundary conditions.  This size was determined to be much larger
than the Debye length of the system.  The Debye length is the screening distance
of a charged particle in a solution, beyond which its interaction with other ions
is effectively screened. The Debye length is given as
\begin{equation}
\label{eq_debyelength}
\lambda_D = {1 \over \sqrt { 8 \pi l_B n_0}}
\end{equation}
where $n_0$ is the molar conentration of ions in the reservoir, that is, the
concentration of the counter ions remaining outside the sphere.
As will be seen in the results, the microgel equilibrates to a radius of $15$,
with an average of $46\%$ of the counter ions remaining outside the micorgel.
This results to a Debye length of $14.1$.  The closest distance between 
two periodic images of the microgel is $54$, which is roughly $3.8$ times
this Debye length.  This method of calculating the molar concentration $n_0$
already yeilds the upper limit of the Debyle length.  Choosing any other method
for defining the $n_0$---for example, $n_0$ as the total number of counter ions
over the total box volume---would yeild a higher value for $n_0$, and thus
a shorter Debye length.  This set-up therefore guarantees enough screening
to prevent the microgel from being strongly electrostatically affected
by its periodic image.  Thus the microgel is in the dilute limit where each
micgrogel behaves independently of other microgels.  This allows us to compare
the results with the PB calculations of a colloid in a cell model, which
also assumes that the interaction of one colloid with other colloids is
negligible.

\subsection{Iterative Poisson-Boltzmann Solution}
\label{sec_pb_description}

The Poisson-Boltzmann (PB) theory is a mean-field theory that describes
the electrostatic interactions between charged particles in solution.
The Poisson-Boltzmann equation is
\begin{equation}
\label{eq_pb}
\nabla^2 \phi \left( r \right) =
- {\frac{4 \pi e_0}{\epsilon_0 \epsilon_r}}\sum_i{ n_i^0 z_i
\exp{ \left[ -e_0 z_i \phi \left( r \right) / k_BT \right] } }
\end{equation}
where $\phi(r)$ represents the electrostatic potential, and $z_i$ is the valency
of the ion species $i$.  The charge density $\rho(r)$, assumed to be a Bolztmann
distribution of charges, is given by
\begin{equation}
\label{eq_chargedist}
\rho \left( r \right) = - e_0 \sum_i{ n_i^0 z_i
\exp{ \left[ -e_0 z_i \phi \left( r \right) / k_BT \right] } }
\end{equation}
This theory assumes that the only interactions
taken into consideration are Coulombic interactions between point charges, thus
neglecting any finite size effects of the charges.  The other assumptions are that
the ion entropy is that of an ideal gas.

In a cell model \cite{Katchalsky_1971}, one can approximate the interactions of a
charged object with its counter ions in a local volume, and assume that the
inteaction of this unit with other similar units is negligible.  Using this model,
we can therefore think of a microgel as one such charged object isolated in a 
spherical volume.  This spherical cell model is shown in in figure \ref{fig_pb_to_mg}a.
The inner circle represents the charged object---for example, a colloid or
a microgel---where the total (positive) charge of the object is distributed
evenly within the sphere.  This charged object has radius $r_0$.  It is cocentric
to an outer spherical cell of radius $R$, as represented by the outer
circle in the figure.  Counter ions are present in this spherical cell.  In 
this study, these counter ions were represented as point charges with a charge
of $-1.0$.  They were allowed to go anywhere within the outer sphere,
even into the charged object.

Our MG set-up (figures \ref{fig_microgel} and \ref{fig_pb_to_mg}e)
is fairly close to a charged sphere inside a spherical cell
model (figure \ref{fig_pb_to_mg}a).  Solving the PB equation for this model
of a charged sphere would therefore be a good comparison
for the resulting distribution of charges of the MG simulation.
The PB equation is a non-linear partial differential equation, and 
has closed-form analytical solutions only for a limited number of cases.
For our case, it has to be solved numerically, as was done by Deserno et. al.
\cite{Deserno_2001,Deserno_etal_2004}, where the PB equation was solved
numerically for the same spherical cell model as described above.

Using this PB numerical solver, we calculated the fraction of counter ions present
in the charged sphere.  The radius of the outer spherical cell was set to $R=52$. 
The inner charged sphere had a radius of $r_0=18$ and was given a total charge
of $+250.0$.  There were $n_{CI}=250$ counter ions present, each with a charge
of $-1.0$.  The Bjerrum length was chosen to be  $l_B=1.0$. These parameters
were chosen  so as to match those of the MG simulation: the volume of the
spherical cell is equal to the volume of the simulation box, and the charged
sphere radius of $r_0=18$ is equal to the energy minimized structure
(figure \ref{fig_pb_to_mg}c) of the microgel coming from the initial
structure (figure \ref{fig_microgel}a).  The parameters and results
of this calculation, labelled as \textbf{PB-1}, are listed in
Table \ref{tab_param_pb_ls_st}.  The results will be discussed in the suceeding
sections.  One other PB calculation was done,
labelled as \textbf{PB-2}, with slightly different parameters listed
in \ref{tab_param_pb_mg}.  The reason for this will be explained in the discussion
of the MG simulation in section \ref{sec_ssmg}.

\subsection{Rationale of the Simulation Strategy}
\label{sec_pb_to_mg}

The charged sphere in a cell model is already quite close to the
microgel simulation.  However, there still are quite a number
of assumptions made between the PB model and the MG simulation.
These are outlined as follows:

\begin{enumerate}
\item \label{a_dp} \textbf{Discrete Particles}.  Whereas the total charge
of the sphere in the PB model is uniformly distributed throughout
the charged sphere, discrete charges are used in the MG simulation.
\item \label{a_ev} \textbf{Excluded Volume}. The microgel particles, whether
charged or uncharged, in the MG simulation have excluded volume, as parameterized
by $\sigma_{LJ}$, which in turn dertermines the local binding energy.
This is non-existent in the charged sphere of the PB solver.
\item \label{a_ap} \textbf{Arranged Particles}.  The charged particles in the
MG simulation are situated along the chain, and are therefore not totally
free to move relative to each other, differing from the even distribution
of charges in the PB model.
\item \label{a_tf} \textbf{Thermal Fluctuations}.  Aside from being arranged
along the chains in the microgel, the charged particles also move as the polymer
itself moves due to thermal fluctuations.  This is not included in the PB theory.
\item \label{a_pbc} \textbf{Periodic Boundary Conditions}.  Lastly, the PB model
was calculated in an isolated cell model, whereas the MG simulations were done under
periodic boundary conditions.  This difference also compares the electrostatic
algorithms employed in the two systems.  When a simulation is done in
a cell model, a direct sum of the Coulombic potential between all pairs
of charged particles is calculated, whereas an Ewald sum approach---in this paper,
the P3M algorithm---is done for the MG simulation in the periodic box.
\end{enumerate}

Due to these assumptions, one would easily suspect that there would also
be differences between the results of the PB calculation and the microgel simulation.
However, it would be impossible to directly link the differences in results
to each of these assumptions, since all of these assumptions are simultaneously
present and could easily confuse the analysis.  We therefore performed
a series of model simulations that systematically remove from the original model
one or more of the assumptions listed, thereby going from the PB picture to
the MG simulation in a step-wise fashion.  Using this procedure, we are able
to attribute specific effects to the corresponding assumptions.  These simulations
are called Lattice Sphere (\textbf{C-LS}), Static Microgel (\textbf{C-ST}),
and Microgel in a Sperical Shell (\textbf{C-MG}), which will be discussed below.
The \textbf{C} in the labels indicate that they are done using
a cell model, as compared to the microgel (\textbf{P-MG}), which was simulated using periodic
boundary conditions (hence, labelled \textbf{P}).  The set-ups of these simulations
are shown in  figure \ref{fig_pb_to_mg}.  The numbers on the arrows in this figure
indicate the assumptions removed when going from one simulation to the next.

In all the calculations and simulations in this paper (as shown in figure \ref{fig_pb_to_mg}),
the following variables were kept constant:

\begin{enumerate}
\item \label{cv_tot_vol} \textbf{Total volume}.  The total volume in which counter ions
where free to move were the same for all simulations and \textbf{PB} calculations.
An outer radius of $R=52.0$ was used for all the calculations involving the 
spherical cell model.  A cube with the same volume (within a $0.63\%$ difference) has
a length of $L=84$, which is the value used for the simulations invovling a cubic box.
\item \label{cv_nCI} \textbf{Counter Ions}. The total number of counter ions
was $n_{CI}=250$, each with a charge of $q_{CI}=-1.0$.
\item \label{cv_qSph} \textbf{Charge of Inner Sphere}.  In order to keep the system
electrically neutral, the total charge of the inner sphere was kept at $q_{sp}=+250.0$.
In some simulations, however, the total number of charged particles in the inner sphere
were greater than $250$.  In these cases, the fractional charge of these particles 
was adjusted in order to keep the total charge at $+250.0$.
\end{enumerate}

\section{Lattice Sphere Simulations (\textbf{C-LS})}
\label{sec_ls}

\subsection{\textbf{C-LS} Methodology}
\label{sec_ls_method}

We simulated a series of systems which we call Lattice Sphere (\textbf{C-LS}),
which consist of charged particles that are fixed on a cubic grid.  The set-up
is shown in figure \ref{fig_pb_to_mg}b. The final shape of this lattice approximates
a sphere. This spherical lattice of radius $r_0=18$ was fixed at the center of
an outer sphere with radius $R=52$.  Counter ions were  placed inside the outer sphere.
The outer sphere is given the same LJ repulsive term with parameters in
Table \ref{tab_sim_common} so as to keep the counter ions inside the sphere.
During the simulation, the counter ions are free to move throughout the whole sphere 
(note that $R=52$ is the radius of the volume available for counter ion motion).
A repulsive Lennard-Jones potential was used between counter ions and
the charged particles in the lattice so as to avoid that two oppositely charged
particles to interact too strongly via Coulombic forces.  No Lennard-Jones potential
was used between counter ions, since these were treated as point particles, following
the assumption of PB theory (section \ref{sec_pb_description}).
This set-up is very similar to the PB model, except that two assumptions were
already were lifted: discrete charged particles (assumption \ref{a_dp}) 
distributed as evenly as possible within a sphere,
and the added excluded volume (assumption \ref{a_ev}) of these particles
instead of an even charge distribution.

All \textbf{C-LS} simulations had a total of $n_{CI}=250$ counter ions each with
a charge of $-1.0$, and a total charge of $+250.0$ for the lattice.  Given the 
various possible cubic lattice arrangements of charged particles with an overall
spherical shape, the three arrangements used in this study had 250, 484 and 894
charged particles in the  lattice sphere.  And, in order for this sphere to retain
a total of +250, charges of +1.0, +0.516, and +0.279, were distributed in these
respective arrangments.  In doing so, the point charges were systematically
smeared out within the sphere, going from 250 (+1.0) to 484 (+0.516) and lastly
to 894 (+0.279), therefore lessening the effect of charge discretization.
In addition, the $\sigma_{LJ}$ was also decreased from $\sigma=1.0$ to smaller values.
This has the effect of minimizing the effect of excluded volume (assumption \ref{a_ev}).

We used two ways of determining these smaller $\sigma_{LJ}$ values.
The first way was to maintain the energy of closest approach, that is,
the Coulombic energy between the counter ion (with a charge of -1.0) and
the charged particle (whether charged +1.0, +0.516, or +0.279) in
the three different systems were always be the same.  By examining
equation \ref{eq_coulomb}, varying both the charge of the particle
($q_i$ in the numerator) and the minimum distance ($r_{ij}$ in the
denominator) in the same way would yield equal energies.  Thus, when varying the the charge
from +1.0 to +0.516 to +0.279, we also varied the $\sigma$ of the LJ potential
in equation \ref{eq_lj} from $\sigma=1.0$ to $\sigma=0.516$ to $\sigma=0.279$.
These are done in three calculations: \textbf{C-LS-1} with $\sigma=1.0$, $q_i=+1.0$, 
and $n_{CP}=250$; \textbf{C-LS-2} with $\sigma=0.516$, $q_i=+0.516$ and $n_{CP}=484$;
and \textbf{C-LS-3} with $\sigma=0.279$, $q_i=+0.279$ and $n_{CP}=894$.
Therefore the differences in results of these three calculations cannot be due to
the energy of closest approach.

The second procedure used in decreasing the values of $\sigma$ is by maintaining
the ratio of the volume occupied by the charged particle to the volume of the entire
inner sphere.  In the previous procedure, the ratio of volumes of the charge particle
to the inner sphere were $1.68 \times 10^{-3}$ for \textbf{C-LS-1}, 
$4.44 \times 10^{-4}$ for \textbf{C-LS-2}, and $1.30 \times 10^{-4}$ for \textbf{C-LS-3}.
Keeping this ratio equal to \textbf{C-LS-1}, the $\sigma_{LJ}$ was changed to the
following values for a new set of calculations: \textbf{C-LS-4} with $\sigma_{LJ}=0.802$,
$q_i=+0.516$ and $n_{CP}=484$; and \textbf{C-LS-5} with $\sigma_{LJ}=0.654$,
$q_i=+0.279$, and  $n_{CP}=894$.  Thus having kept the free volume inside
the inner sphere constant, we can safely assume that the differences
in results of these three calculations would not be due to entropic effects, which
have been significantly reduced in this procedure.

These five Lattice Sphere simulations aimed at isolating the effects of the first two
PB assumptions. The parameters and results are summarized in table \ref{tab_param_pb_ls_st}.
These simulations were all done in a spherical cell model, with the
electrostatics potential computed as a direct sum of Coulombic energies
between all pairs of particles.

\subsection{Discussion: Result of Charge Discretization and Excluded Volume}
\label{sec_dp}

Refering to table \ref{tab_param_pb_ls_st}, the first discussion compares the
results of \textbf{PB-1}, \textbf{C-LS-1}, \textbf{C-LS-2}, and \textbf{C-LS-3}.
These four calculations have the following parameters that are the same:
total charge of sphere, number of counter ions, radius of charged sphere/lattice,
and outer radius.  From calculations \textbf{C-LS-1} to \textbf{C-LS-2}
to \textbf{C-LS-3}, the total number of discrete charged particles in the inner sphere
was increased from 250 to 484 to 894, with a corresponding
decrease in partial charge and $\sigma_{LJ}$ from 1.0 to 0.517 to 0.279.
Figure \ref{fig_intprob} shows the integrated counter ion probability
of the simulations listed in Table \ref{tab_param_pb_ls_st}.  The second
to the last line in Table \ref{tab_param_pb_ls_st} refer to the results of the
percentage counter ion in figure \ref{fig_intprob} where $r$ is equal to
the sphere radius $r_0$. The dots in the inset of
figure \ref{fig_intprob} at $r_0=18 \sigma$ indicate these percentages.
Figure \ref{fig_densities} shows the density of the counter ions
as a function of the distance from the center of the inner sphere.

In general, the integrated counter ion probability (Fig. \ref{fig_intprob})
and the counter ion density (Fig. \ref{fig_densities}) graphs
of the \textbf{C-LS} simulations are close to the \textbf{PB-1} graph,
with \textbf{C-LS-1} being the farthest and lower and \textbf{PB-1},
\textbf{C-LS-2} being closer but higher than \textbf{PB-1},
and \textbf{C-LS-3} being practically the same as the \textbf{PB-1}.
The fraction of counter ions inside the sphere at $r_0=18$
(last line of Table \ref{tab_param_pb_ls_st} and the inset in Figure \ref{fig_intprob})
also reflects this trend.  As the number of charged particles is increased
(with the corresponding decrease in partial charge) and thus the model
becoming closer to the PB picture, the result of counter ion distribution
approaches the PB calculation. Of the three LS calculations,
the one with a unit charge of $+1.0$ can be considered closest of real
physical systems, since charged particles always have unit charges.
Thus we can say that the discretization of charges in a real system
would have a counter ion condensation around $3\%$ lower than what is
predicted by PB theory.

However, there seems to be a slight anomaly in the trend shown above.  Although
a stepwise smearing out of the charges should have corresponded to a gradual
approach to the \textbf{PB-1} calculation ($52.83\%$), the results showed
a jump from the lower value for \textbf{C-LS-1} ($49.80\%$), to higher for
\textbf{C-LS-2} ($53.85\%$), then back down to practically equal for
\textbf{C-LS-3} ($52.78\%$)  to the \textbf{PB-1} calculation.
This trend cannot be caused by the energy of closest approach,
since this has been set-up to be equal in these three
LS calculations.  We then proceed to examining the second assumption applied to the
LS calculations, that of the excluded volume of the charged particles.
The three calculations whose exculded volume, as measured by the ratio of volumes
of charge particle to that of the inner sphere, are \textbf{C-LS-1}, \textbf{C-LS-4},
and \textbf{C-LS-5}.  The integrated counter ion distribution and the counter ion
densities of \textbf{C-LS-2} and \textbf{C-LS-4} are practically the same.  The same is
true for the graphs of \textbf{C-LS-3} and \textbf{C-LS-5}.  The percentage of
counter ions inside the inner sphere are listed at the bottom of
table \ref{tab_param_pb_ls_st}.  The trends are the same as the first set of
LS calculations: \textbf{C-LS-1} is lower ($49.80\%$) than
\textbf{PB-1} ($52.83\%$), \textbf{C-LS-4} is higher
($53.35\%$), and \textbf{C-LS-5} ($52.23\%$) is closest to
\textbf{PB-1}.  This shows that the counter ion distributions is not dependent
on the excluded volume of the charged particles.

The explanation of this trend simply comes from the ground state energies of the
different set-ups, since the ground-state structures depend mainly on the
configuration of the system.  To do this, we set the Langevin temperature $T=0.0$
and using a small friction coefficient $\Gamma$ between 0.01 to 0.001 to allow for
slow relaxation to the ground state.  This resulting trend of the ground state
electrostatic energies of these systems are similar
to the trends in percentage counter ion in the  sphere:
\textbf{C-LS-1} being the highest $-2160.72 k_B T$ (least stable),
\textbf{C-LS-2} being the highest $-2301.29 k_B T$ (most stable), and
\textbf{C-LS-3} in the middle $-2293.17 k_B T$.  Thus the trend in counter ion
fraction in the lattice sphere depends mostly on ground state configurations
rather than the excluded volume of the charged particles. This does not mean,
however, that excluded volume will not have an effect in polymeric systems.
The next section discusses the more realistic arrangement of charged particles
within a polymer chain.

\section{Static Microgel Simulation Details (\textbf{C-ST})}
\label{sec_stmg}

\subsection{\textbf{C-ST} Methodology}

After having seen the effects of assumptions \ref{a_dp} and \ref{a_ev},
we then proceeded to simulate another system with assumption \ref{a_ap} removed,
that of charged particles in the system being placed along the chains in the microgel.
In this case, we took the coordinates of an expanded microgel
with the same parameters as described in \ref{sec_mg_sim_details} and
fixed their coordinates so as not to allow thermal fluctuations.  A snapshot of
this system is shown in Figure \ref{fig_pb_to_mg}c.  This simulation is labelled
as \textbf{C-ST}.  Two versions of this simulation were performed.  In the first
case (\textbf{C-ST-1}), we removed the uncharged monomers (total of 230)
so as not to include the effects of their excluded volume
in the simulation.  In the second case (\textbf{C-ST-2}), the uncharged monomers
were not removed, so that the microgel has the same number of charged and uncharged
particles present in the microgel simulations (\textbf{C-MG} and \textbf{P-MG}).
These two therefore tested the effect of the arranging the 
charged particles along a chain, with the second looking at the added effect
of the excluded volume of the uncharged particles, which is therefore one
step closer to the MG set-up.  The parameters are similar
to \textbf{PB} and \textbf{C-LS}, as listed in Table \ref{tab_param_pb_ls_st}.
This particular arrangement of the particles
had an overall effect of lessening the space between the
charged particles beside each other on  the same chain, and increasing
empty spaces between charged particles on different chains.
As with the Lattice Sphere case, this simulation was also done in a
spherical cell model, with the electrostatics calculated as a direct sum.

\subsection{Discussion: Arrangement of Excluded Volume along a Polymer}
\label{sec_arr}

The parameters and results of the static microgel simulations,
\textbf{C-ST-1} and \textbf{C-ST-2}, are listed in Table \ref{tab_param_pb_ls_st},
with the integrated counter ion probability curve shown in figure \ref{fig_intprob}
and the counter ion density shown in figure \ref{fig_densities}.
The results show a further decrease of the fraction of counter ions in
the static microgel \textbf{C-ST-1} ($46.28\%$) and in \textbf{C-ST-2} ($45.67\%$)
as compared to \textbf{PB-1} ($52.83\%$) and \textbf{C-LS-1} ($49.80\%$).
Figure \ref{fig_exclvol} shows the  relative distances of particles for the
calculations of: a) \textbf{C-LS-1}, b) \textbf{C-ST-1}, and c) \textbf{C-ST-2}.
The black circles represent the charged particles in the lattice sphere
or microgel, the grey circle in fig. \ref{fig_exclvol}c represents
the uncharged particle in \textbf{C-ST-2}, the dotted circles represent
the excluded volume of these particles $\sigma_{LJ}$,
and the unshaded circles represent the counter ions.

In the case of the \textbf{C-LS-1} configurations (fig. \ref{fig_exclvol}a),
there is enough space in between two charged particles for a counter ions
to move around the charged particle.  The closest distance between the two
dotted circles (given by $x$) is greater than $1.0$, which is the
Bjerrum length in the simulation. Thus even in this case, their Coulombic
repulsion would still be much less than their thermal energy.

That is not the case for the static microgel (\textbf{C-ST-1}, fig. \ref{fig_exclvol}b)
which does not include the uncharged particles in the microgel.
The closest distance between paths (dotted circles) of two counter ions of adjacent
charged particles is less than the Bjerrum length.  Thus, unlike the lattice sphere
cases, the space right in between two charged particles cannot be occupied
by two counter ions, thus lessening the overall free space available for
the counter ions.  This effect could in turn decrease the fraction of
charges in the microgel.  The results of the second static microgel simulation
(\textbf{C-ST-2}, fig. \ref{fig_exclvol}c) further supports this point.
In this case, the additional uncharged particle (grey circle)
contributes the extra excluded volume to the same space where
electrostatic interactions prevented the counter ions to go, as argued in the
case of \textbf{C-ST-1}.  Unlike the $3.52\%$ drop from \textbf{C-LS-1}
to \textbf{C-ST-1}, these two static microgel simulations have very close
percentages of counter ions inside the gel---$46.28\%$ for \textbf{C-ST-1}
and $45.67\%$ for \textbf{C-ST-2}---indicating that the additional uncharged particles
contributed only slightly to the reduction of counter ions in the gel.
Thus the proximity of the charged particles in \textbf{C-ST-1}
arranged along the polymers in the static microgel provided enough counter ion
repulsion to lessen the avaialble space for counter ion motion and thus decrease
the overall counter ion percentage in the sphere going from the \textbf{C-LS} to the
\textbf{C-ST} calculations.

Although excluded volume had no effect on the configuration of the lattice sphere
simulations as described in section \ref{sec_dp}, the specific arrangement of the
excluded volume, in this case, arranged in a polymer chain, can slightly lessen the
available space for counter ion condensation and thus lower the fraction of
counter ions inside the microgel.

\section{Microgel in a Spherical Cell (\textbf{C-MG}) and in a Periodic Box (\textbf{P-MG})}
\label{sec_ssmg}

\subsection{\textbf{C-MG} and \textbf{P-MG} Methodology}

After removing assumptions \ref{a_dp} and \ref{a_ev} to the lattice sphere
(\textbf{C-LS}) and assumption \ref{a_ap} to the static micro gel (\textbf{C-ST}),
we then removed the next assumption \ref{a_tf}, which looks at the effect of
thermal fluctuations of the charged particles in the polymer.
This is basically a microgel simulation, where the particles of the microgel
are allowed to move, inside a spherical cell.
Only the central particle of the microgel was kept fixed so as to keep
the microgel in the center of the sphere.  The parameters are basically the same
as those of the microgel, and are listed in Table \ref{tab_param_pb_ls_st}.
Aside from thermal fluctuations, the other difference of between this system and
static microgel in a sphere (section \ref{sec_stmg}) is the decrease in microgel
radius, thus decreasing the free volume in between the chains.  The parameters are
listed in Table \ref{tab_param_pb_ls_st}.  This simulation is labeled \textbf{C-MG}.
As with the \textbf{C-LS} and \textbf{C-ST} simulations, the \textbf{C-MG}
simulation was also done in a spherical cell model, with the electrostatics
calculated as a direct sum.

To determine the effect of the last assumtion, that of periodic boundary conditions
and differences in electrostatic algorithm, we compared the results of 
\textbf{C-MG} to \textbf{P-MG}.  The methodology and parameters for the \textbf{P-MG}
simulation has already been discussed in section \ref{sec_mg_sim_details}.

Due to the thermal fluctuation of the entire microgel for both \textbf{C-MG} and
\textbf{P-MG}, the radius of the microgel was also fluctuating, but generally
maintaing its overall spherical shape.  In order to determine the average radius
of the microgel, we used the following equation\cite{Rubenstein_Colby_2003} that
relates the square radius of gyration of a rigid sphere to the radius of the sphere
\begin{equation}
\label{eq_rg2_r}
R_g^2 = { 3 \over 5 } r^2
\end{equation}
where $R_g^2$ is square radius of gyration of the sphere with radius $r$.

\subsection{Discussion: Thermal Fluctuations and Periodic Boundary Conditions}
\label{sec_pb_mg}

The parameters and results (second to the last line) of the \textbf{C-MG}
and \textbf{P-MG}  are listed in Table \ref{tab_param_pb_mg}.  The integrated
counter ion probability is shown in figure \ref{fig_intprob} and the counter ion
density is shown in figure \ref{fig_densities}.  Allowing the microgel particles
to move allowed the microgel shrink from $r_0=18$ for the static microgel
simulation (\textbf{C-ST-1}) to an average of $r_0=14.96$ for \textbf{C-MG}
and $r_0=14.99$ for \textbf{P-MG}. However, the fraction of counter ions
within these radii for each simulation increased to $54.03\%$ for \textbf{C-MG}
and $53.95\%$ for \textbf{P-MG}.  Due to the difference of radius between \textbf{PB-1}
and these two microgel simulations, another PB calculation,\textbf{PB-2} was done
with a radius of $r_0=14.96$.  The corresponding fraction of counter ions in the
inner sphere for \textbf{PB-2} also increased to $55.59\%$, resulting in a decrease 
of $2.6\%$ when going from PB calculation to MG simulation of the same radius $r_0$.
As already seen in the discussion of the LS results in section \ref{sec_dp},
the simluation results correspond to the PB calculation, with a slight decrease
of a similar value ($3.0\%$) in the fraction of counter ions in the lattice sphere.  
This shows that the thermal fluctuations of charged particles in a polymer
do not affect the counter ion fraction inside the MG.

The decrease of the radius of the charged microgel/sphere leads to an increase
in fraction of counter ions in the micogel/sphere. This can be explained as follows.
The decrease of the microgel/sphere radius, while keeping the total charge constant,
corresponds to an increase in the overall charge density within the sphere, thus
attracting more counter ions into the sphere.  Our results of the PB calculations
show that the gain in energy outweighs the loss of entropy, which the confined
confined counter ions experience.

Both \textbf{C-MG} and \textbf{P-MG} have practically the same results---configuration
as measured by $R^2_g$, fraction of counter ion inside the microgel, integrated counter
ion probability (firgure \ref{fig_intprob}) and counter ion densities
(figure \ref{fig_densities}).  Thus the spherical cell model simulation, whose
electrostatics is calculated as a direct sum, is equivalent to the corresponding
simulation under periodic conditions, as long as the sytsems are set up correctly.
This means that the system size is done ensuring enough electrostatic screening (as
calculated by the Debye length) in order to satisfy the ``isolated system''
assumption of the cell model, and that the P3M electrostatics is tuned correctly.

\section{Conclusions}
\label{sec_conc}

We have simulated a model polyelectrolyte microgel setup with the corresponding
counter ions in a periodic box, using the results mainly to calculate the
fraction of counter ions inside the microgel.  We compared these results to a
Poisson-Boltzmann numerical solution\cite{Deserno_2001,Deserno_etal_2004}
in order to see how the mean field solution differs from the results
of the simulation.  We outlined the differences
between these two models, listed as assumptions in section \ref{sec_pb_to_mg},
and proceeded to perform a series of simulations aimed at removing these assumptions
in a step-wise fashion in order to isolate the effects of each assumption.

The very first assumption of going from a charged sphere with uniform distribution
in the PB solver to having discrete charges showed a slight decrease in the
fraction of counter ions inside the microgel.  The fact that the sucessive ``smearing out''
of the discrete charges, thus approacing the uniform charge distribution
in the PB solver, resluted in a sucessive approach of the LS results to that of the
PB calculation shows that the discretization of charges causes the slight decrease
of the fraction of counter ions inside the microgel.  A further decrease is shown
when going from the lattice sphere with evenly spaced charge distribution
within the sphere to the static microgel configuration, where the charges are not as
evenly distributed since they were lined up within the polymer.
Aside from the discretization of charges, the density of charges of the microgel
particles also plays a role in fraction of counter ions in the microgel, the two
being directly proportional.

Even with the many assumptions in between the MG simulations and PB theory,
we have seen that the results of the two methods are quite close to one another.
We can therefore safely assume that if PB models with configurations that follow the
more complex configurations found in molecular simulations of real systems are constructed,
the results in terms of counter ion condensation from both approaches would be
comparatively close, as long as we stick the the basic assumptions used in this study
such as movable charges and counter ions, good solvent, and relatively isolated systems.
This allows one to use the PB cell model to predict structural or thermodynamic properties
of microgels in solution.

\section*{Acknowledgments}

We are grateful to DFG for the funding under the Schwerpunktprogramm SPP 1259
``Intelligente Hydrogele''.  We would like to thank Markus Deserno for allowing
us to use his PB numerical slover program.
Gil Claudio would also like to thank Torsten St\"uhn, Olaf Lenz, and
Vagelis Harmandaris for their technical support.


\clearpage



\begin{table}
\caption{Parameters common to all simulations in this paper.  These are:
         Lattice Sphere(\textbf{C-LS}), Static Microgel (\textbf{C-ST}),
         Microgel in a Spherical Cell (\textbf{C-MG}), and Microgel in a
         Periodic Box (\textbf{P-MG}).}
\label{tab_sim_common}
\begin{center}
\begin{tabular}{lc}
\hline
Parameter                               & value    \\ \hline
$k_BT$                                  & 1.0      \\
$\epsilon_{LJ}$                         & 1.0      \\
$l_B$                                   & 1.0      \\
unit mass $m$                           & 1.0      \\
time step                               & 0.012    \\
ensemble                                & NVT      \\
thermostat                              & Langevin \\
friction coefficient ($\Gamma$)         & 1.0      \\
time steps for every snapshot           & 1000     \\
minimum number of snapshots for results & 15000    \\
program used                            & ESPResSo \\
\hline
\end{tabular}
\end{center}
\end{table}

\clearpage

\begin{table}
\caption{Simulation parameters for the calculations discussed in sections \ref{sec_dp}
         and \ref{sec_arr}.  The following parameters are common for all these
         calculations: the radius of the inner sphere $r_0=18 \sigma$ whose total charge
         is $q_{sp}=+250.0$, the total number of counter ions is $n_{CI}=250$,
         the charge for each coutner ion is $q_{CI}=-1.0$,
         each counter ion is treated as a point particle with no $\sigma_{LJ}$
         interaction between them, all use the cell model and the electrostatic interactions
         are calculated via direct sum, the outer radius of the cell is $R=52\sigma$.
         The entries left blank in this table are parameters that are not applicable
         to the specific computation.}
\label{tab_param_pb_ls_st}
\begin{center}
\begin{tabular}{l|cccc|cc|cc}
\hline
Simulation label                           & \textbf{PB-1}  & \textbf{C-LS-1} & \textbf{C-LS-2} & \textbf{C-LS-3} & \textbf{C-LS-4} & \textbf{C-LS-5} & \textbf{C-ST-1} & \textbf{C-ST-2} \\ \hline
number of charged particles ($CP$)         &       & 250    & 484    & 894    & 484    & 894    & 250    & 250    \\
number of uncharged particles              &       & 0      & 0      & 0      & 0      & 0      & 9      & 239    \\
excluded volume ($\sigma_{LJ}$) of $CP$    &       & 1.0    & 0.516  & 0.279  & 0.802  & 0.654  & 1.0    & 1.0    \\
charge of particle ($q_{CP}$)              &       & 1.0    & 0.516  & 0.279  & 0.516  & 0.279  & 1.0    & 1.0    \\
\hline
percentage $CI$ in microgel/sphere         & 52.83 & 49.80  & 53.85  & 52.78  & 53.36  & 52.21  & 46.28  & 45.67  \\
error of percentage $CI$                   &       & 2.01   & 1.99   & 2.00   & 2.00   & 1.98   & 2.12   & 2.09   \\
\hline
\end{tabular}
\end{center}
\end{table}

\clearpage

\begin{table}
\caption{Simulation parameters for the calculations discussed in section \ref{sec_pb_mg}.
         The following parameters are common for all these calculations:
         the total charge of each particle in the inner sphere is $q_{sp}=+250.0$,
         the total number of counter ions is $n_{CI}=250$, the charge of each counter ion is
         $q_{CI}=-1.0$, the electrostatics of \textbf{C-MG} was calculated
         via direct sum, while the electrostatics of \textbf{P-MG} calculation which
         uses periodic boundary conditions was calculated using P3M.
         The entries left blank in this table are parameters and results that are not
         applicable to the specific computation.}
\label{tab_param_pb_mg}
\begin{center}
\begin{tabular}{l|cccc}
\hline
Simulation label                           & \textbf{PB-2} & \textbf{C-MG} & \textbf{P-MG} \\ \hline
radius of sphere $r_0$ (constant)          & 14.96 &        &        \\
boundary (Cell or Periodic box)            & cell  & cell   & box    \\
box length $L$ (in $\sigma$)               &       &        & 84     \\
outer radius $R$ (in $\sigma$)             & 52    & 52     &        \\
number of charged particles ($CP$)         &       & 250    & 250    \\
number of uncharged particles              &       & 239    & 239    \\
excluded volume ($\sigma_{LJ}$) of $CP$    &       & 1.0    & 1.0    \\
charge of particle ($q_{CP}$)              &       & 1.0    & 1.0    \\
excluded volume of ($\sigma_{LJ}$) of $CI$ &       &        & 1.0    \\
\hline
average squared radius of gyration $R_g^2$ &       & 134.27 & 134.76 \\
error in $R_g^2$                           &       & 4.62   & 4.51   \\
radius of microgel (eqn. \ref{eq_rg2_r})   &       & 14.96  & 14.99  \\
percentage $CI$ in microgel/sphere         & 55.59 & 54.03  & 53.95  \\
error in percentage $CI$                   &       & 2.36   & 2.32   \\
\hline
\end{tabular}
\end{center}
\end{table}


\clearpage

\begin{figure}
\begin{center}
\includegraphics[scale=0.75]{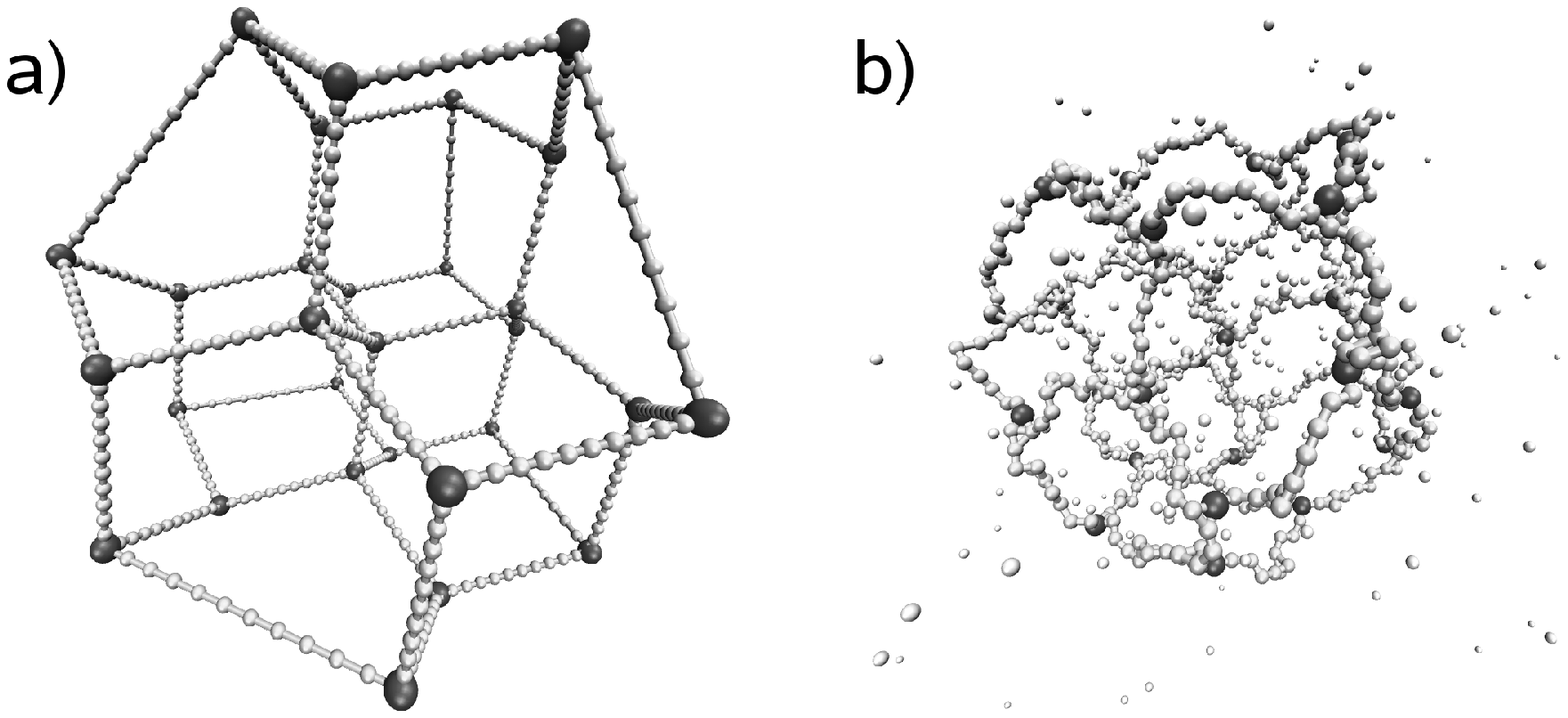}
\caption{}
\label{fig_microgel}
\end{center}
\end{figure}

\noindent
\textbf{Figure \ref{fig_microgel}}.
Snapshot of a microgel \textbf{P-MG} simulation (shown in perspective) in a
periodic box.  Figure a. shows how the starting structure was constructed,
while figure b. shows one equilibrated structure.  The microgel is composed
of 46 polymers,  each composed of 10 monomers (represented as grey shaded beads).
The polymer also contains an additional 29 beads acting as crosslinks (represented
as dark beads, larger radius is done only for emphasis), which connect the ends
of either three or four polymers.  Thus the microgel has a total of 489 beads.
Half of the monomers and 20 crosslinks, totalling 250 beads, have a +1.0 charge,
giving a total charge of +250.0.  The box contains 250 counter ions (not all shown),
each with a charge of -1.0, making the box have an overall neutral charge.

\vspace{12 pt}

\begin{figure}
\begin{center}
\includegraphics[scale=0.5]{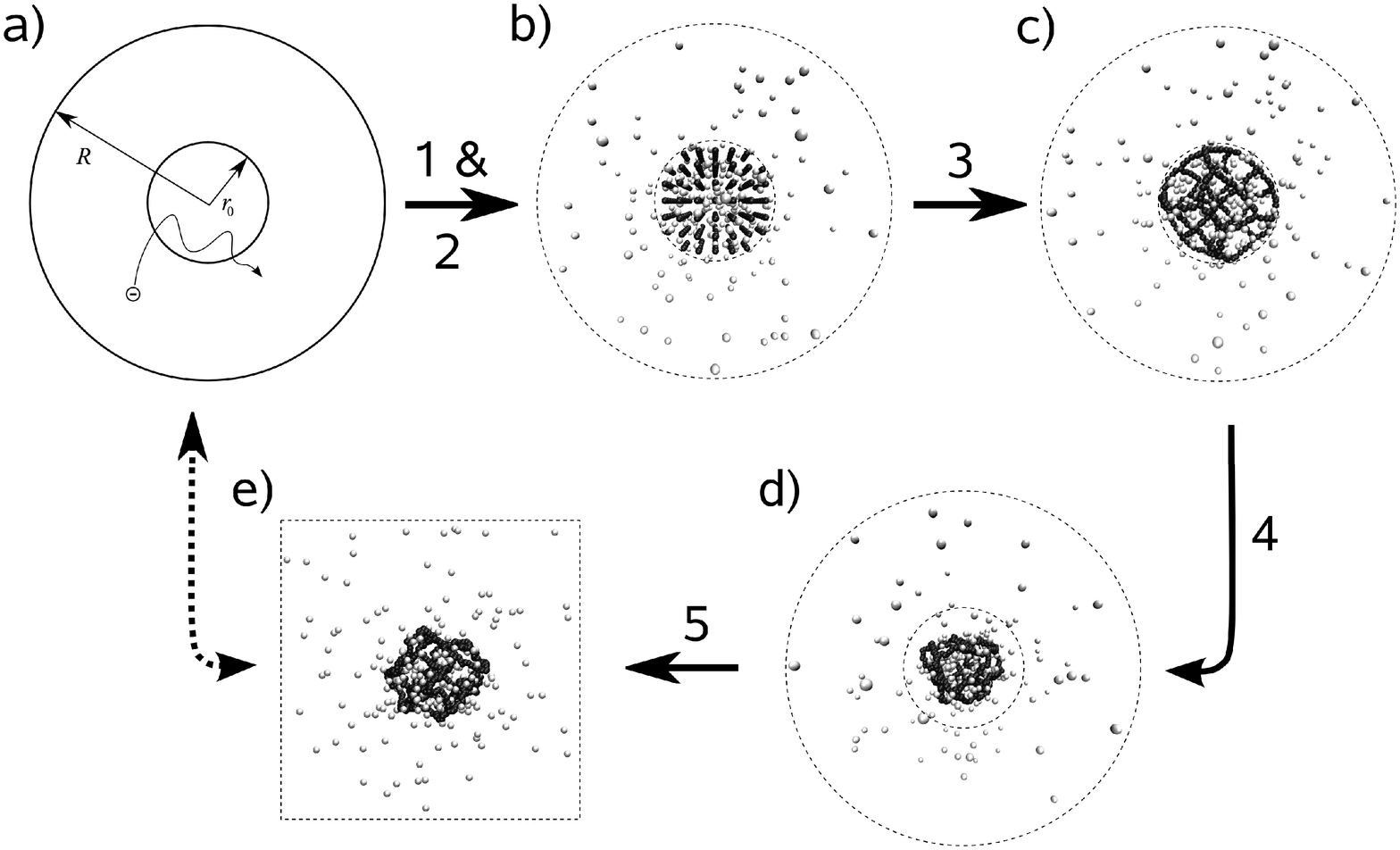}
\caption{}
\label{fig_pb_to_mg}
\end{center}
\end{figure}

\noindent
\textbf{Figure \ref{fig_pb_to_mg}}.
Models discussed in section \ref{sec_pb_to_mg} to go from the Poisson-Boltzmann
spherical cell model (\textbf{PB}) to the microgel simulation (\textbf{P-MG}).
The dark objects are the particles in the lattice (b) or the monomers in the microgel
(c-e), while the lighter colored objects are the counterions that are free to move within
the boundaries.  The numbers on the arrows indicate the assumptions taken when going
from one model to the next, as discussed in section \ref{sec_pb_to_mg}.
a. Poisson-Boltzann spherical cell model (\textbf{PB}). b. Lattice sphere (\textbf{C-LS}).
c. Static microgel (\textbf{C-ST}). d. Cell microgel (\textbf{C-MG}). e. Microgel
(\textbf{P-MG}).

\vspace{12 pt}

\begin{figure}
\begin{center}
\includegraphics[scale=0.75]{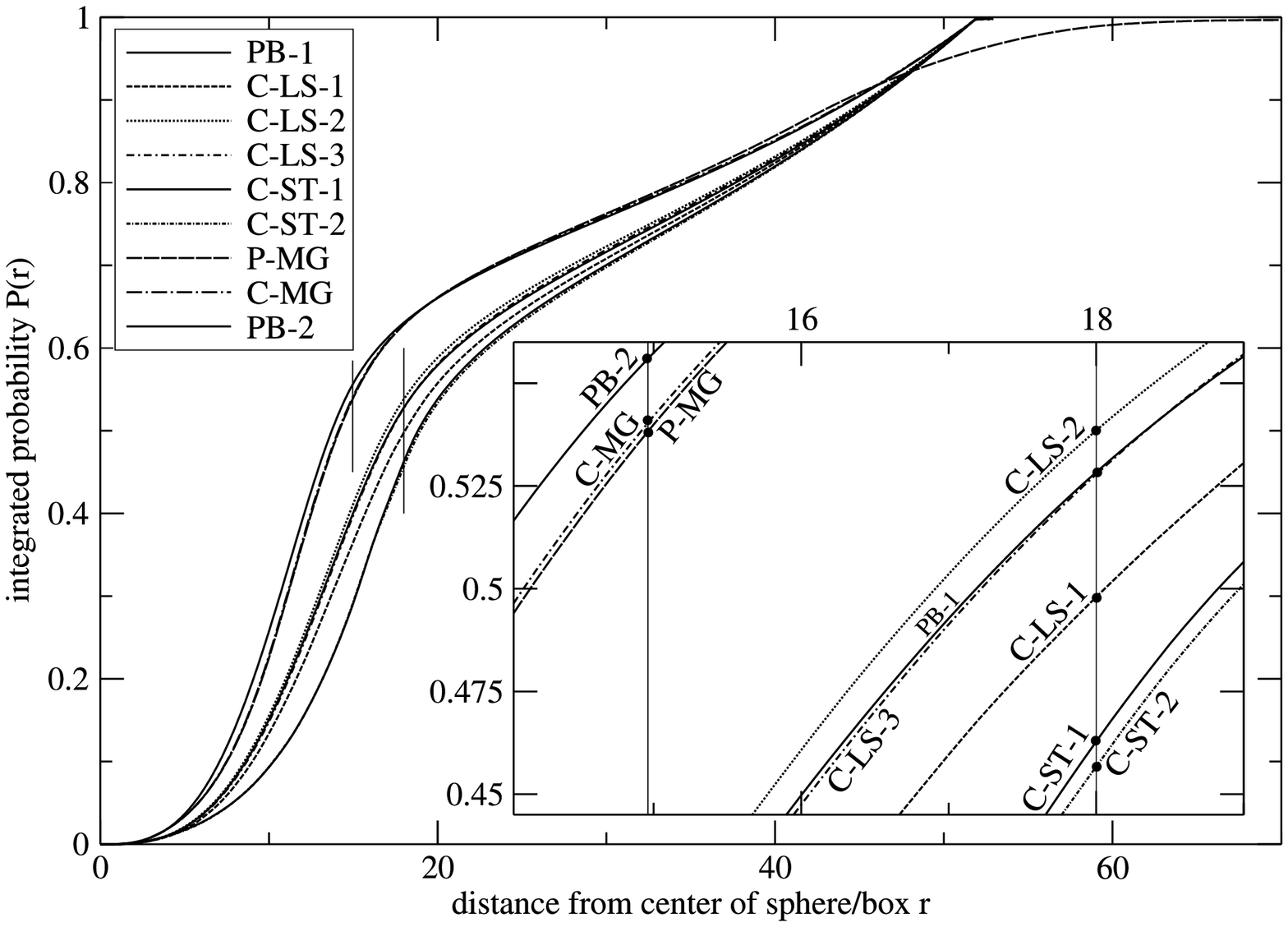}
\caption{}
\label{fig_intprob}
\end{center}
\end{figure}

\begin{figure}
\begin{center}
\includegraphics[scale=0.75]{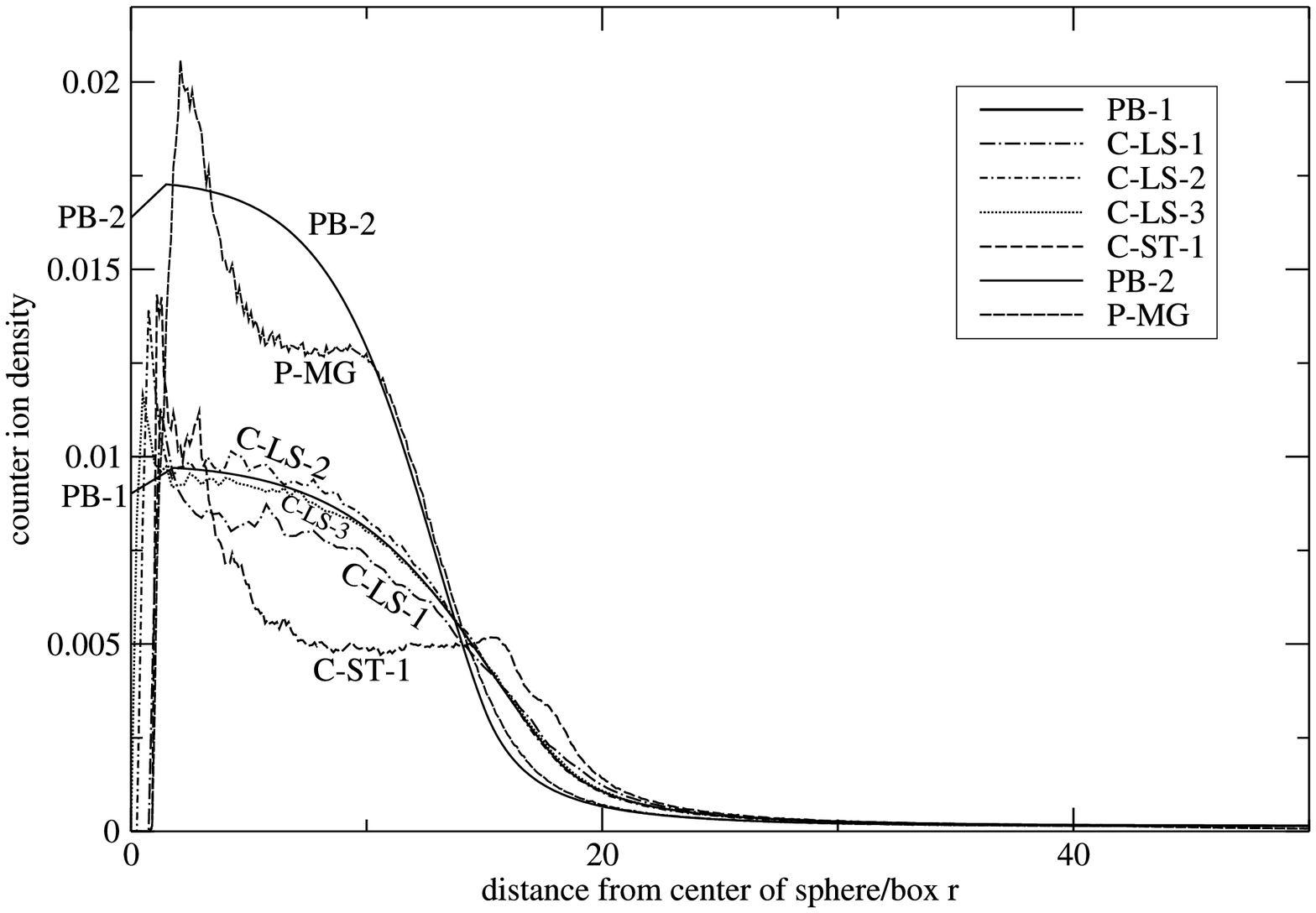}
\caption{}
\label{fig_densities}
\end{center}
\end{figure}

\noindent
\textbf{Figure \ref{fig_intprob}}.
Integrated ion probability of simulations listed in Tables \ref{tab_param_pb_ls_st}
and \ref{tab_param_pb_mg}, also corresponding to the counter ion densities in figure
\ref{fig_densities}.  The inset is a magnification of the regions around $r_0=18$
(radius corresponding to table \ref{tab_param_pb_ls_st}) and around $r_0=14.99$
(radius corresponding to table \ref{tab_param_pb_mg}).  The dots in the inset
indicate the numbers reported at the last line in Tables \ref{tab_param_pb_ls_st}
and \ref{tab_param_pb_mg}.

\vspace{12 pt}

\noindent
\textbf{Figure \ref{fig_densities}}.
Counter ion densities of the simulations listed in Tables \ref{tab_param_pb_ls_st} and
\ref{tab_param_pb_mg}, also corresponding to the integrated counter ion probabilities
in figure \ref{fig_intprob}.  The counter ion density graph of \textbf{C-LS-4}
is very simlar to \textbf{C-LS-2}, that of \textbf{C-LS-5} to \textbf{C-LS-3},
that of \textbf{C-ST-2} to \textbf{C-ST-1}, and that of \textbf{C-MG} to \textbf{P-MG}.
These are therefore not included in this figure.

\vspace{12 pt}

\begin{figure}
\begin{center}
\includegraphics[scale=0.75]{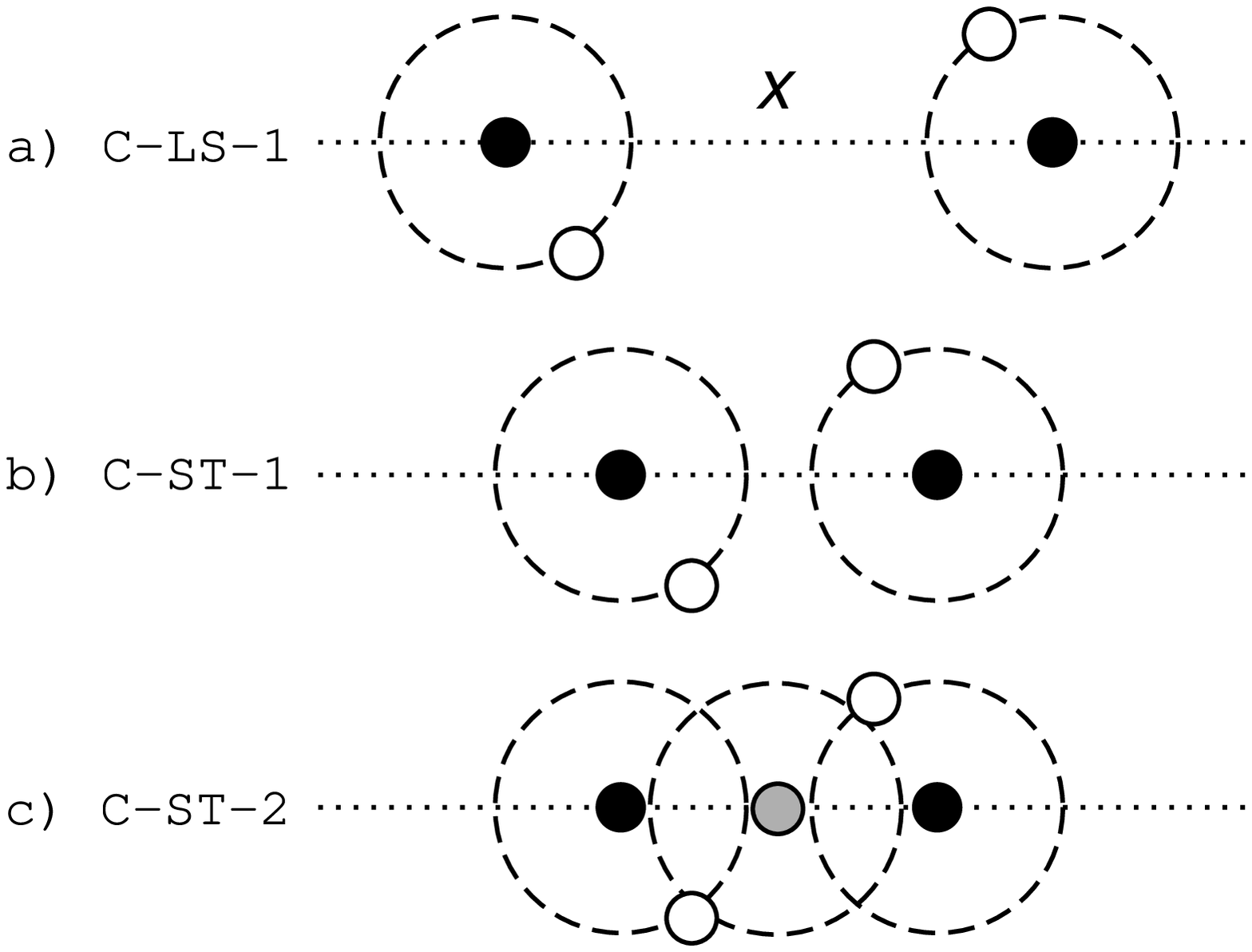}
\caption{}
\label{fig_exclvol}
\end{center}
\end{figure}

\noindent
\textbf{Figure \ref{fig_exclvol}}.
Arrangement of Excluded Volume. Solid circles represent the charged particles, open
circles the counter ions, the grey shaded circle represents an uncharged particle,
and the dashed circles the excluded volume $\sigma_{LJ}$ between the particles
and the counter ions.  a. Lattice sphere \textbf{C-LS-1}.  The charged
particles are at a distance of $4.81$ apart for the \textbf{C-LS-1} case,
$3.67$ for the \textbf{C-LS-2} case, and $3.04$ for the \textbf{C-LS-3}
case.  After subtracting the two excluded volume radii from the charged particle,
the distance $x$ is still greater than Bjerrum length $l_B$, thus allowing the counter ion
to move freely around the dashed circle. b. Static microgel \textbf{C-ST-1} without
uncharged particles.  Since the distance between charged particles is only around
$2.0$, there is no extra space for the counter ions to freely move once the
excluded volume has been taken into account. c. Static microgel \textbf{C-ST-2}
with uncharged particles in between.  The decrease in free space in b and c results
in the lowering of the counter ion fraction for the ST case as compared to the LS.


\clearpage

\clearpage

\clearpage

\clearpage

\clearpage

\end{document}